\documentclass[conference,a4paper]{IEEEtran}
\IEEEoverridecommandlockouts
\usepackage[binary-units,per-mode=symbol,per-symbol=p]{siunitx}
\usepackage{cite}
\usepackage{adjustbox}
\usepackage[ruled,vlined]{algorithm2e}
\usepackage{amsmath,amssymb,amsfonts}
\usepackage{algorithmic}
\usepackage{graphicx}
\usepackage{textcomp}
\usepackage[colorlinks=false, pdfborder={0 0 0}, breaklinks]{hyperref}
\usepackage{xcolor}
\usepackage{balance}
\usepackage{enumitem}
\setlist{nolistsep,leftmargin=*}
\usepackage{setspace}
\usepackage{wrapfig}
\usepackage{listings}
\usepackage{color}
\usepackage{caption}
\usepackage{subcaption}
\usepackage{todonotes}
\usepackage[longtable]{multirow}

\def\BibTeX{{\rm B\kern-.05em{\sc i\kern-.025em b}\kern-.08em
    T\kern-.1667em\lower.7ex\hbox{E}\kern-.125emX}}

\usepackage{amsmath}

\usepackage{ctable}

\newcommand{\cf}{\emph{cf.}\xspace}

\newcommand{\rev}[1]{\textcolor{black}{#1}}
\newcommand{\amr}[1]{\textcolor{black}{#1}}

\newcommand{\EY}{$E_{\text{Y}}$\xspace}
\newcommand{\LY}{$L_{\text{Y}}$\xspace}

\newcommand{\h}{$h$\xspace}

\newcommand{\BDRP}{$BDR_{\text{P}}$}
\newcommand{\BDRV}{$BDR_{\text{V}}$}
\newcommand{\BDRH}{$BDR_{\text{H}}$}

\begin{document}

\title{Rate-Quality-Energy Convex Hull for High Dynamic Range Adaptive Video Streaming}


\author{\IEEEauthorblockN{Amritha Premkumar\IEEEauthorrefmark{1}, Christian Herglotz\IEEEauthorrefmark{1}
}
\IEEEauthorblockA{\IEEEauthorrefmark{1}Chair of Computer Engineering, Brandenburg University of Technology Cottbus-Senftenberg, Germany}
}

\maketitle

\begin{abstract}
High Dynamic Range (HDR) video introduces substantially higher decoding complexity and energy consumption than Standard Dynamic Range (SDR), yet most adaptive streaming pipelines still rely on bitrate ladders optimized using SDR‑oriented metrics such as VMAF. These ladders fail to capture HDR‑specific perceptual behavior and ignore decoder‑side energy, resulting in energetically inefficient operating points. This work proposes a \rev{rate–quality–energy convex‑hull optimization framework} that jointly considers bitrate, \mbox{HDR‑aware} perceptual quality (HDRVMAF), and \rev{client‑side processing energy}, including decoding and upscaling. Experiments on twelve 4K HDR sequences from the LIVE HDR Video Quality Assessment Database evaluate 72 representations per clip and demonstrate that traditional VMAF‑optimized ladders frequently select high‑resolution encodes that become dominated once energy is accounted for. Compared to the \emph{Default} HLS ladder, our HDR‑aware method achieves substantial bitrate savings of up to \SI{54.34}{\percent} at equal HDRVMAF, while yielding \rev{relative client‑side energy reductions of up to} \SI{43.10}{\percent} under higher energy weighting. 
These results \rev{demonstrate the effectiveness of the proposed framework as a practical foundation for constructing HDR bitrate ladders that explicitly expose and control perceptual \amr{quality}–energy trade‑offs}.
\end{abstract}


\section{Introduction}
\begin{figure}[t]
\centering
\begin{subfigure}{0.475\columnwidth}
\centering
    \includegraphics[width=\columnwidth]{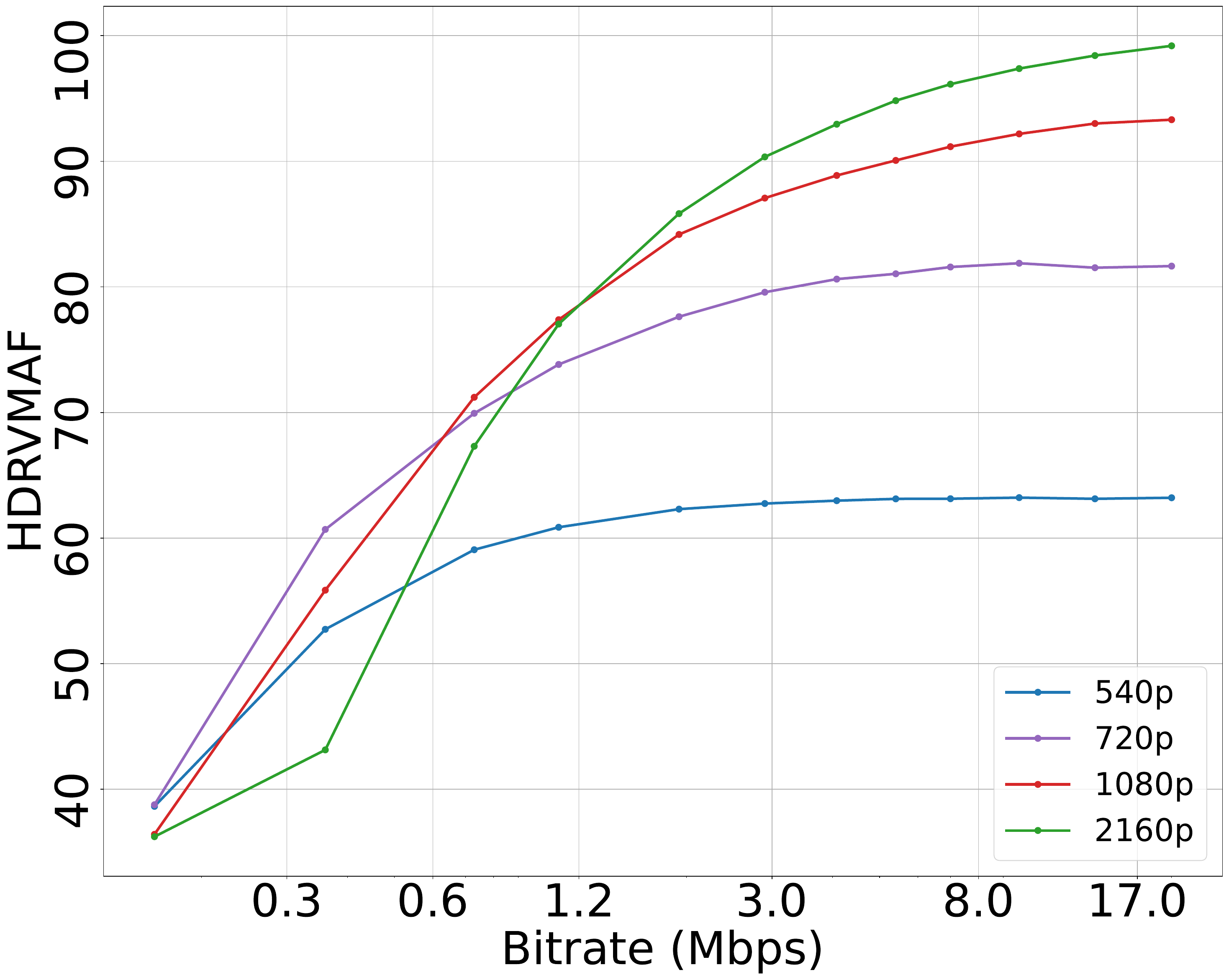}
    \includegraphics[width=\columnwidth]{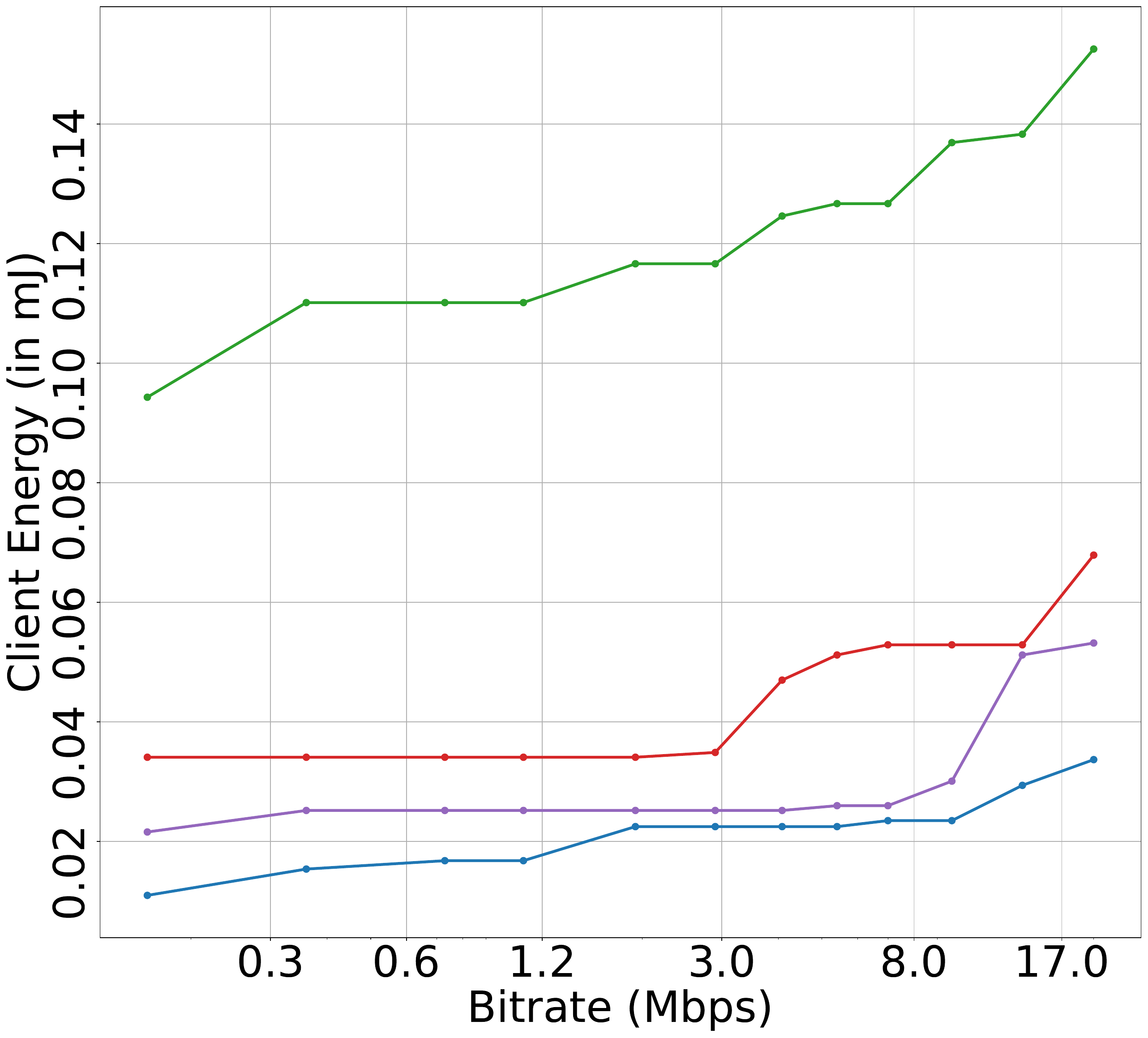} 
\caption{\emph{CourtYard}}
\end{subfigure}
\begin{subfigure}{0.475\columnwidth}
\centering
    \includegraphics[width=\columnwidth]{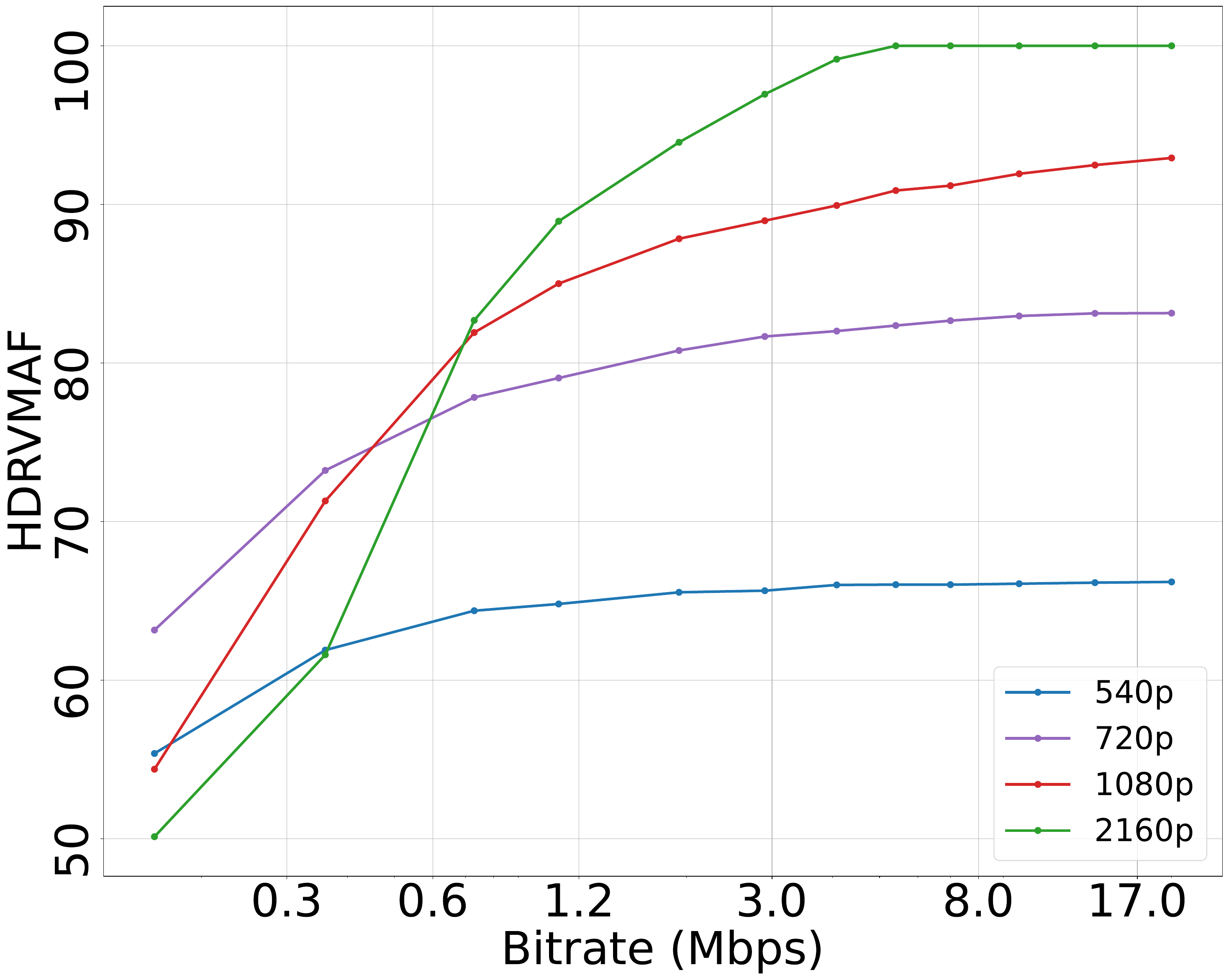}
    \includegraphics[width=\columnwidth]{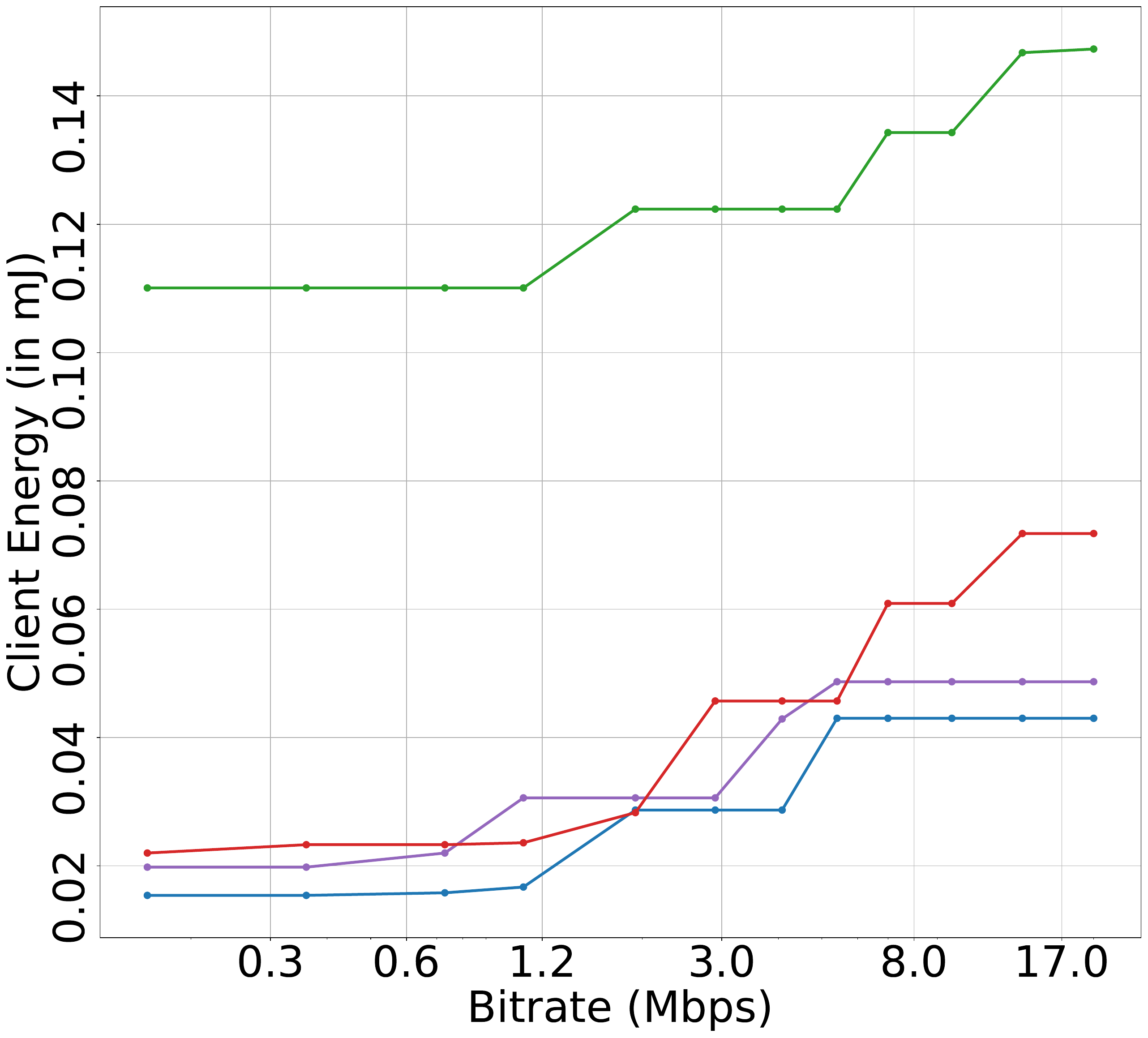} 
\caption{\emph{NightStreet}}
\end{subfigure}
\caption{Rate-quality, and rate-energy curves for the representative HDR sequences of LIVE HDR Video Quality Assessment Database for various resolutions.}
\vspace{-0.9em}
\label{fig:rd_intro}
\end{figure}

High Dynamic Range (HDR) video has become a key component of modern streaming services due to its extended luminance range, higher peak brightness, and wider color gamut compared with Standard Dynamic Range (SDR) content. These characteristics improve visual realism but also increase encoding and decoding complexity, leading to higher energy consumption on client devices, particularly mobile platforms and smart TVs with power-constrained system-on-chip (SoC) architectures~\cite{herglotz_energy_2024}. Consequently, HDR streaming systems must optimize not only perceptual quality but also playback energy efficiency. \amr{Fig.~\ref{fig:rd_intro} provides an illustrative example of the relationship between bitrate, HDR perceptual quality, and client-side processing energy for representative HDR content.} Higher resolutions achieve superior HDRVMAF~\cite{hdrmax} scores but require substantially greater client-side processing energy, whereas lower resolutions reduce energy consumption at the expense of perceptual quality.

Adaptive streaming addresses bandwidth variations by encoding content at multiple resolutions and bitrates, allowing the client to select an appropriate representation. Conventional bitrate-ladder construction relies on rate--quality convex hulls generated using SDR-oriented metrics such as VMAF~\cite{blog_vmaf_2018}. Although effective for compression efficiency, these methods assume that bitrate and perceptual quality alone determine optimal operating points. In practice, different representations impose different computational costs during playback. High-resolution and high-bit-depth videos require more intensive decoding because of larger pixel counts, higher memory bandwidth, and more complex prediction structures~\cite{HEVC, mercat_hdr}. Conversely, lower-resolution streams require spatial upscaling, which also incurs non-negligible energy. These effects are particularly pronounced for HDR video because higher bit depths and HDR-specific codec tools further increase decoder workload. 
\amr{Therefore, bitrate-ladder construction should jointly consider perceptual quality and client-side processing energy, which varies across encoded representations.}

\begin{figure*}[t]
    \centering
    \includegraphics[width=0.87\textwidth]{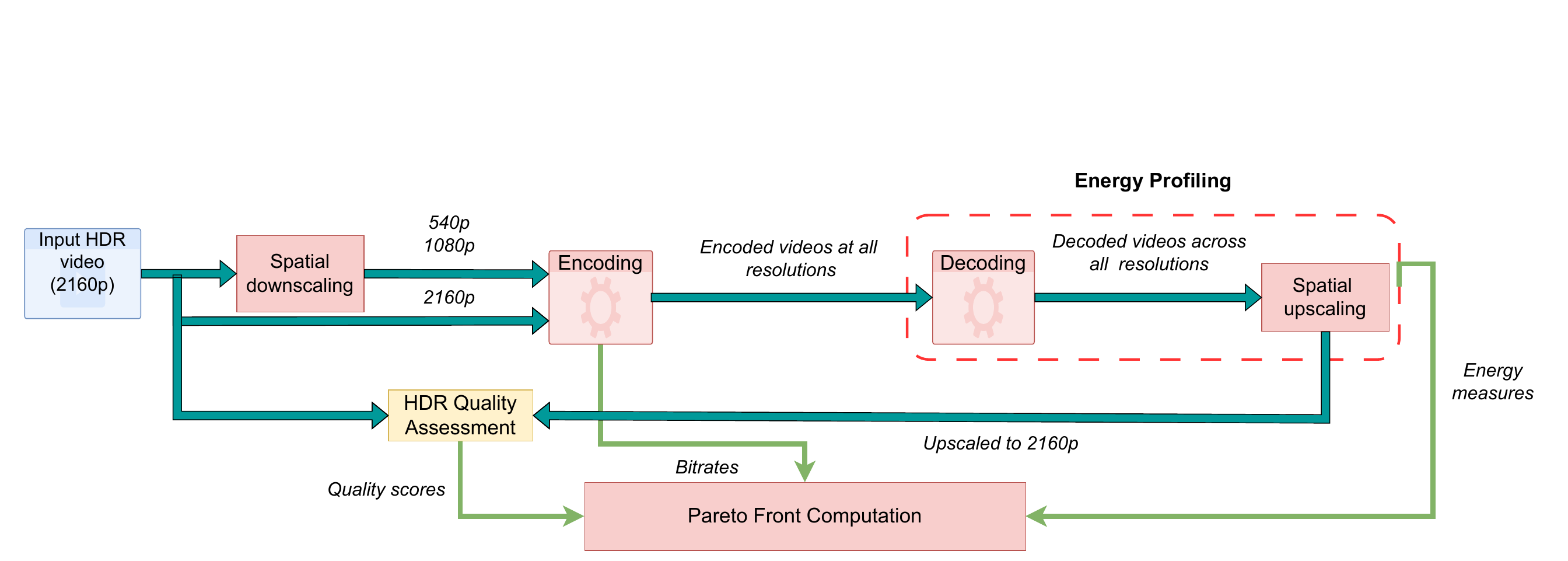}
    \caption{Overview of the proposed rate-quality-energy optimization framework for HDR streaming.}
    \label{fig:system_overview}
\end{figure*}

Existing work on per-title and content-adaptive encoding has demonstrated significant gains in compression efficiency through multi-resolution representation design and convex-hull pruning. Industrial approaches such as Netflix's per-title optimization~\cite{de_cock_complexity-based_2016} and academic methods using PSNR, SSIM, and VMAF~\cite{katsenou_content-gnostic_2019, menon_jnd-aware_2024, Vibhoothi_hdr_av1} have shown that content-dependent ladder construction improves rate--quality efficiency. However, these approaches generally assume uniform decoder cost across representations and ignore the energy implications of resolution adaptation. Energy-aware streaming research has investigated encoder-side energy optimization~\cite{ladre, herglotz_energy_2024} or decoder complexity as a proxy for power consumption~\cite{azimi_eupsico, Rajendran_decodra}. \amr{While these studies provide valuable insights, they either optimize encoder-side processing or use decoder complexity as an indirect indicator of playback cost. However, they rarely incorporate measured client-side processing energy and perceptual quality jointly during bitrate-ladder construction, nor do they account for the energy implications of spatial upscaling.}

Another important challenge is HDR-specific perceptual quality assessment. Traditional metrics such as PSNR and SSIM do not adequately model HDR luminance adaptation, brightness masking, or perceptual thresholds~\cite{Shang_subjective_hdr}. Several HDR-oriented metrics, including HDR Visual-Difference-Predictor (HDR-VDP)-3~\cite{mantiuk2023hdrvdp3}, HDR Video Quality Measure (HDR-VQM)~\cite{hdrvqm}, Fusion of Unified Quality Evaluators (FUNQUE+)~\cite{Venkataramanan_funque_hdr_vqa}, HDRPatchMAX~\cite{Ebenezer_hdr_sdr}, and Display-Independent Gradient Magnitude Similarity (DIGMS)~\cite{zhang_qa_hdr_image}, achieve better agreement with subjective perception. Among these, HDRVMAF~\cite{hdrmax} has emerged as a strong full-reference metric and is adopted as the primary quality measure in this work. 

\amr{Collectively, these observations reveal a gap in current HDR streaming research: bitrate-ladder construction remains primarily driven by rate-quality considerations, while client-side processing energy is typically treated separately. Consequently, the interaction between HDR perceptual quality, decoding energy, and upscaling cost is not explicitly considered during representation selection. }
\amr{To address this gap, we propose a \emph{rate-quality-energy convex-hull} framework for HDR adaptive streaming. The proposed method integrates HDR-aware perceptual quality and client-side processing energy into a unified representation-selection process.}

The main contributions of this work are as follows:

\begin{itemize}[leftmargin=*]
    \item We introduce a rate-quality-energy optimization framework that jointly considers bitrate, HDR perceptual quality, decoder energy, and upscaling energy during bitrate-ladder construction.
    \item We formulate a tunable utility-based representation-selection strategy that exposes and controls the tradeoff between perceptual quality and client-side processing energy.
    \item We demonstrate that many representations selected by conventional quality-driven optimization become suboptimal once client-side energy is considered, and we quantify the resulting bitrate and energy savings.
\end{itemize}    

\section{HDR Rate-Quality-Energy Modeling}
\label{sec:rqemodel}

This section presents the proposed HDR rate-quality-energy modeling framework for representation selection. Each candidate representation is evaluated using three attributes: (i) delivered bitrate, (ii) HDR perceptual quality, and (iii) client-side processing energy, including decoding and optional upscaling. These jointly define the optimization space shown in Fig.~\ref{fig:system_overview}. Unlike conventional rate--distortion models that optimize only bitrate and quality, the proposed formulation explicitly incorporates playback energy, enabling adaptation to both visual requirements and device-level constraints.

\subsection{Representation Evaluation}

A representation is defined by its spatial resolution $r \in \mathcal{R}$ and target bitrate $b \in \mathcal{B}$. Each representation is decoded on the client and, when $r$ is below the display resolution, upscaled before quality assessment. \amr{The upscaling operation is treated generically and may be implemented using conventional interpolation filters or learned video super-resolution methods, depending on the target playback system.} For every $(r,b)$ pair, the framework records:

\begin{itemize}[leftmargin=*]
    \item Bitrate $R(r,b)$: the achieved coded bitrate.
    \item HDR perceptual quality $Q(r,b)$: measured using HDRVMAF by comparing the decoded (and, if necessary, upscaled) output with the HDR reference.
    \item Client-side processing energy $E_{\mathrm{tot}}(r,b)$: the \amr{absolute} energy required for decoding and spatial upscaling.
\end{itemize}

Evaluating these quantities jointly captures tradeoffs between compression efficiency, perceptual quality, and playback energy that are overlooked by conventional rate--quality analysis.

\subsection{Energy Decomposition}

The total client-side processing energy is modeled as
\begin{equation}
E_{\mathrm{tot}}(r,b)=E_{\mathrm{dec}}(r,b)+E_{\mathrm{up}}(r,b),
\label{eq:Etot}
\end{equation}
where $E_{\mathrm{dec}}$ represents codec decoding operations, while $E_{\mathrm{up}}$ denotes the energy associated with spatial upscaling when the encoded resolution is below the display resolution.

This decomposition separates codec complexity from post-processing overhead, enabling the framework to identify whether playback energy is dominated by decoding or scaling.

\subsection{Scalar Utility for Energy--Quality Tradeoff}

To balance perceptual quality and energy efficiency, a scalar utility function is defined as
\begin{equation}
J(r,b;\alpha_D)=Q(r,b)-\alpha_DE_{\mathrm{tot}}(r,b),
\label{eq:utility}
\end{equation}
where $\alpha_D\ge0$ controls the quality--energy tradeoff. Small values prioritize visual quality, whereas larger values increasingly penalize energy consumption. Rather than representing a physical constant, $\alpha_D$ is \amr{a tradeoff parameter (equivalently, a Lagrange multiplier in the constrained formulation of the quality-maximization problem subject to an energy budget).}

To improve comparability across different content and hardware platforms, the utility may also be expressed in normalized form:
\begin{equation}
J'(r,b;\alpha_D)=
\frac{Q(r,b)-Q_{\min}}{Q_{\max}-Q_{\min}}
-\alpha_D
\frac{E_{\mathrm{tot}}(r,b)-E_{\min}}{E_{\max}-E_{\min}},
\label{eq:utility_norm}
\end{equation}
where $(Q_{\min},Q_{\max})$ and $(E_{\min},E_{\max})$ are computed over the candidate representation set.

\subsection{Resolution Selection via Utility Maximization}

For each target bitrate, the optimal spatial resolution is selected by maximizing the utility:
\begin{equation}
r^*(b;\alpha_D)=
\arg\max_{r\in\mathcal{R}} J(r,b;\alpha_D),
\label{eq:select_r}
\end{equation}
or equivalently the normalized utility in Eq.~\eqref{eq:utility_norm}. The resulting bitrate-to-resolution mapping forms an energy-aware bitrate ladder that jointly considers HDR perceptual quality and client-side processing energy.

Because the utility depends on both content characteristics and $\alpha_D$, the selected resolution adapts naturally to different operating conditions. High-detail content favors higher resolutions when quality is prioritized, whereas increasing $\alpha_D$ shifts selection toward lower-energy representations, enabling device-aware and content-adaptive bitrate ladders.

\section{Experimental Setup}

\emph{Dataset:}
The evaluation uses twelve 4K HDR video sequences from the LIVE HDR Video Quality Assessment Database~\cite{}, selected to represent diverse spatial textures, motion characteristics, and luminance distributions typical of modern HDR streaming content. The dataset includes both low- and high-motion scenes, as well as bright and dark sequences, ensuring broad content diversity. All videos are processed at their native 2160p resolution while preserving the HDR format and metadata throughout the encoding, decoding, and quality-assessment pipeline.

\emph{Encoding configuration:}
Each source sequence is encoded at multiple spatial resolutions and bitrates listed in Table~\ref{tab:exp_par}. All encodes use x265 \amr{v4.0 encoder} with the \emph{slower} preset, a ten-bit internal representation and output. \amr{Keeping the codec configuration, GOP structure, and rate-control settings constant minimizes confounding factors, so observed differences in perceptual quality and client-side processing energy are primarily due to resolution and bitrate.}

\begin{table}[t]
\caption{Experimental parameters and values.}
\centering
\resizebox{0.95\columnwidth}{!}{
\begin{tabular}{l||c|c|c|c|c|c}
\specialrule{.12em}{.05em}{.05em}
\specialrule{.12em}{.05em}{.05em}
\emph{Parameter} & \multicolumn{6}{c}{\emph{Values}}\\
\specialrule{.12em}{.05em}{.05em}
\specialrule{.12em}{.05em}{.05em}
$\mathcal{R}$ & \multicolumn{6}{c}{\{ 360, 540, 720, 1080, 1440, 2160 \} } \\
\hline
$\mathcal{B}$ & 0.160 & 0.360 & 0.600 & 0.900 & 1.600 & 2.400 \\
\cmidrule{2-7}
              & 3.400 & 4.500 & 5.800 & 8.100 & 11.600 & 16.800 \\
\hline
$\alpha_D$ & \multicolumn{6}{c}{ [0, 1] }\\
\hline
Target encoder & \multicolumn{6}{c}{x265 [slower][intra period=\SI{1}{\second}]} \\
\hline
Target decoder & \multicolumn{6}{c}{FFmpeg native HEVC decoder} \\
\specialrule{.12em}{.05em}{.05em}
\specialrule{.12em}{.05em}{.05em}
\end{tabular}
}
\vspace{-0.98em}
\label{tab:exp_par}
\end{table}

\emph{Perceptual quality evaluation:}
Perceptual quality is evaluated using PSNR, VMAF~\cite{blog_vmaf_2018}, and HDRVMAF~\cite{hdrmax}, with HDRVMAF serving as the primary metric because of its improved sensitivity to HDR-specific distortions. Each representation is decoded using \amr{FFmpeg} and, when required, upscaled to 4K using bicubic interpolation in FFmpeg before comparison with the original HDR reference. Although commercial devices typically use hardware-accelerated decoding and scaling, the software-based pipeline provides a consistent environment for relative comparison across representations. HDRVMAF offers a more perceptually meaningful assessment than SDR-oriented metrics by better capturing HDR brightness, contrast, and detail.

\emph{Energy measurement:}
Client-side processing energy is measured on an Intel Core i5-1135G7 system with 8\,GB RAM. 
\amr{The decoding (and upscaling) energy is measured} five times, and the reported value is averaged across runs. Energy is measured using CodeCarbon~\cite{codecarbon}, which accesses Intel RAPL counters to estimate CPU and DRAM energy. Background services are minimized, while display and network energy are excluded. 
For the normalized utility in Eq.~\eqref{eq:utility_norm}, fixed quality bounds of $Q_{\min}=0$ and $Q_{\max}=100$ are used, while measured energy values are used directly since the optimization relies on relative dominance rather than absolute energy units.

\emph{Evaluation metrics:}
Bitrate-ladder performance is evaluated using Bjøntegaard-Delta metrics~\cite{DCC_BJDelta}. Bjøntegaard Delta Rate values (\BDRP, \BDRV, and \BDRH) measure the average percentage bitrate difference required to achieve equivalent quality under PSNR, VMAF, and HDRVMAF, respectively. In addition, Bjøntegaard Delta Client Energy (BDCE) quantifies the average relative change in client-side processing energy required to maintain the same HDRVMAF compared with the baseline ladder. Negative values indicate improvements over the reference method.

\emph{Benchmark methods:}
The proposed framework is compared against two widely used bitrate-ladder construction strategies:
\begin{enumerate}
    \item \emph{Default:} The standard Apple HTTP Live Streaming (HLS) bitrate ladder, which uses predefined resolution--bitrate pairs and serves as the baseline representation set~\cite{HLS_ladder_ref}.
    \item \emph{DynResVMAF:} A content-adaptive approach that selects, for each bitrate, the resolution achieving the highest VMAF after decoding and upscaling, thereby providing a quality-oriented upper bound without a fixed ladder constraint~\cite{de_cock_complexity-based_2016,menon_jnd-aware_2024}.
\end{enumerate}

\section{Results and Analysis}

\subsubsection{Rate-Quality-Energy Pareto Fronts}
\begin{figure}[t]
\centering
\begin{subfigure}{0.48\columnwidth}
\centering
    \includegraphics[width=\columnwidth]{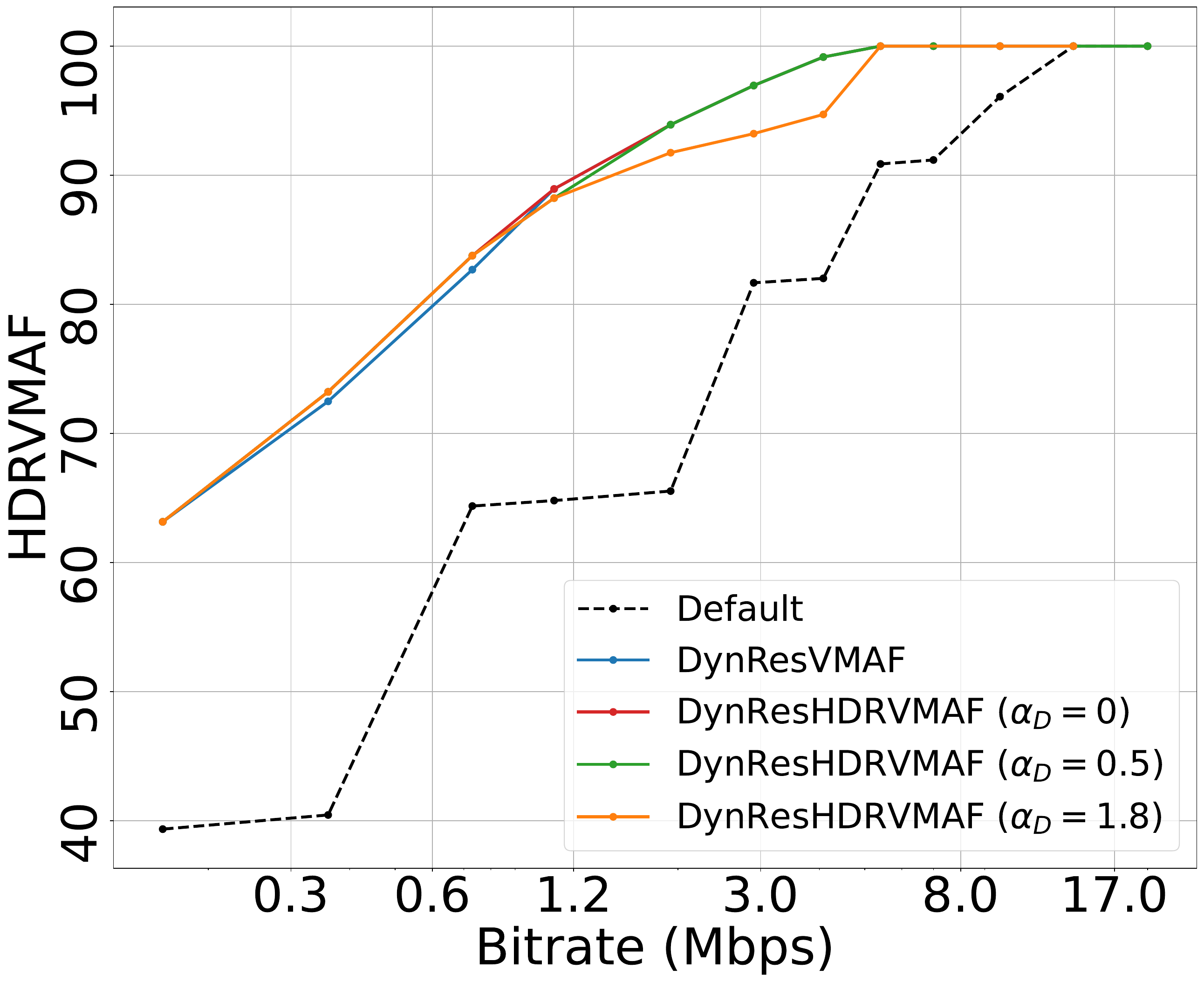}
    \includegraphics[width=\columnwidth]{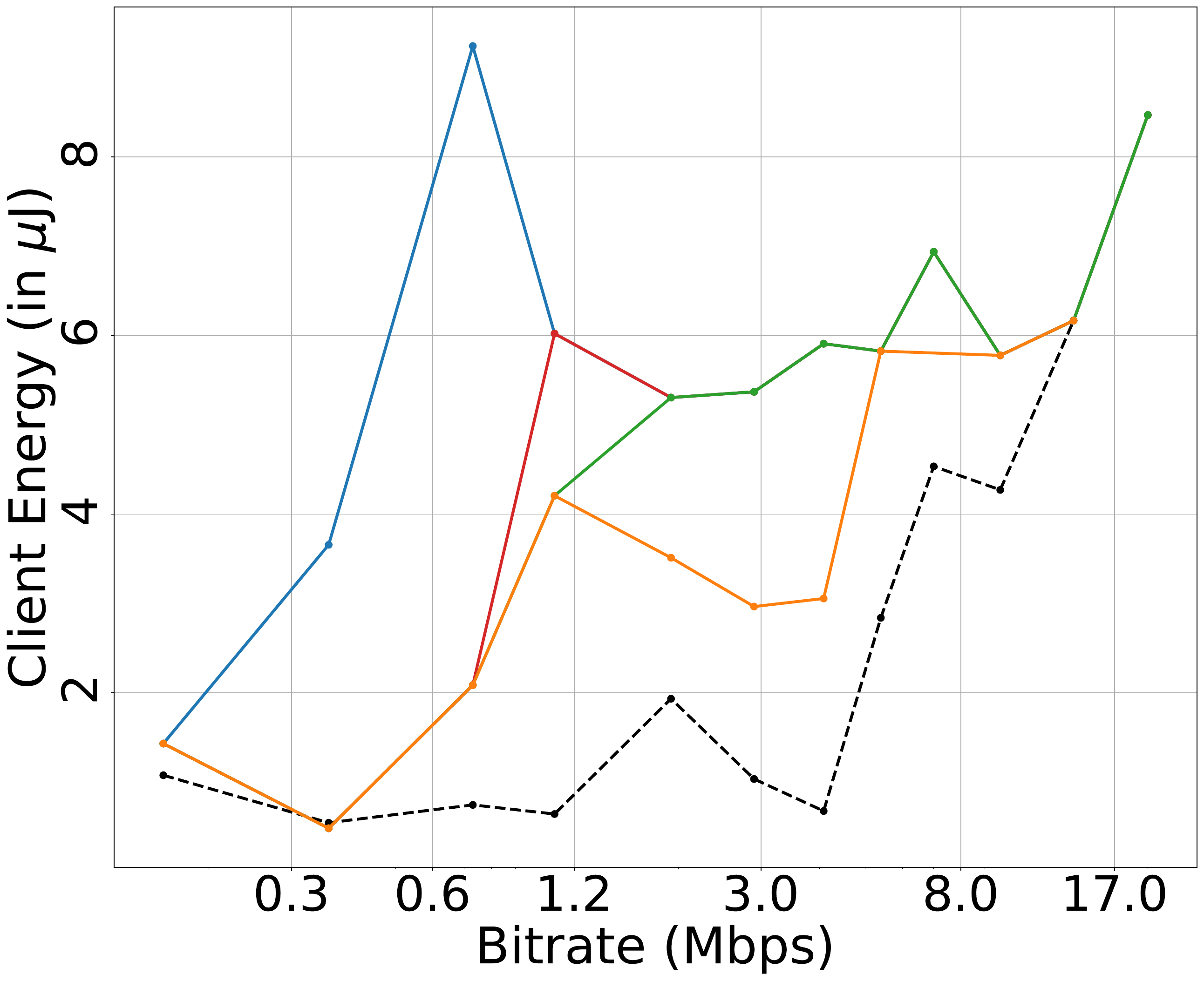} 
\caption{\emph{NightStreet} (\EY:54.99, \h:10.23, \LY:81.49)}
\end{subfigure}
\begin{subfigure}{0.48\columnwidth}
\centering
    \includegraphics[width=\columnwidth]{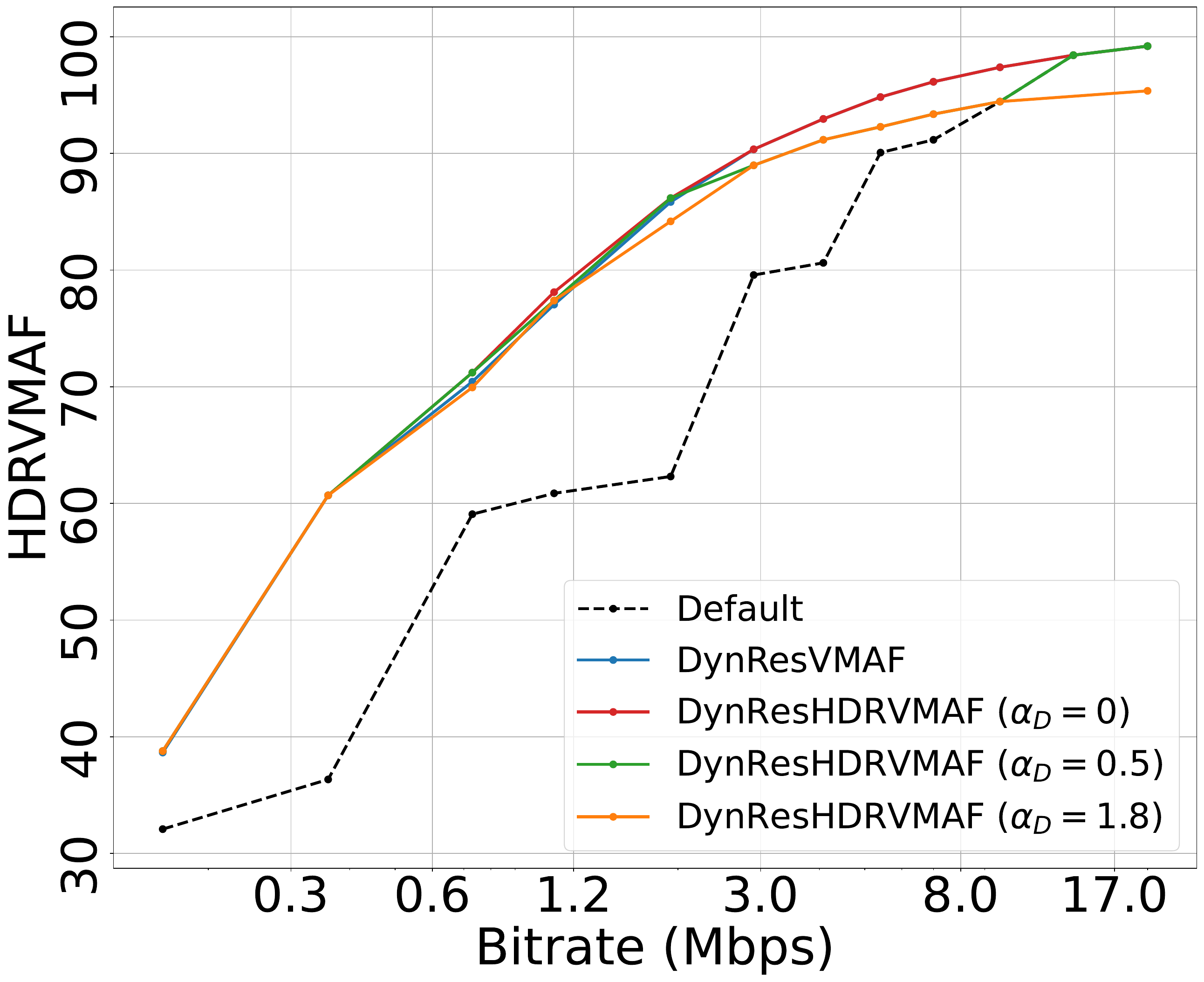}
    \includegraphics[width=\columnwidth]{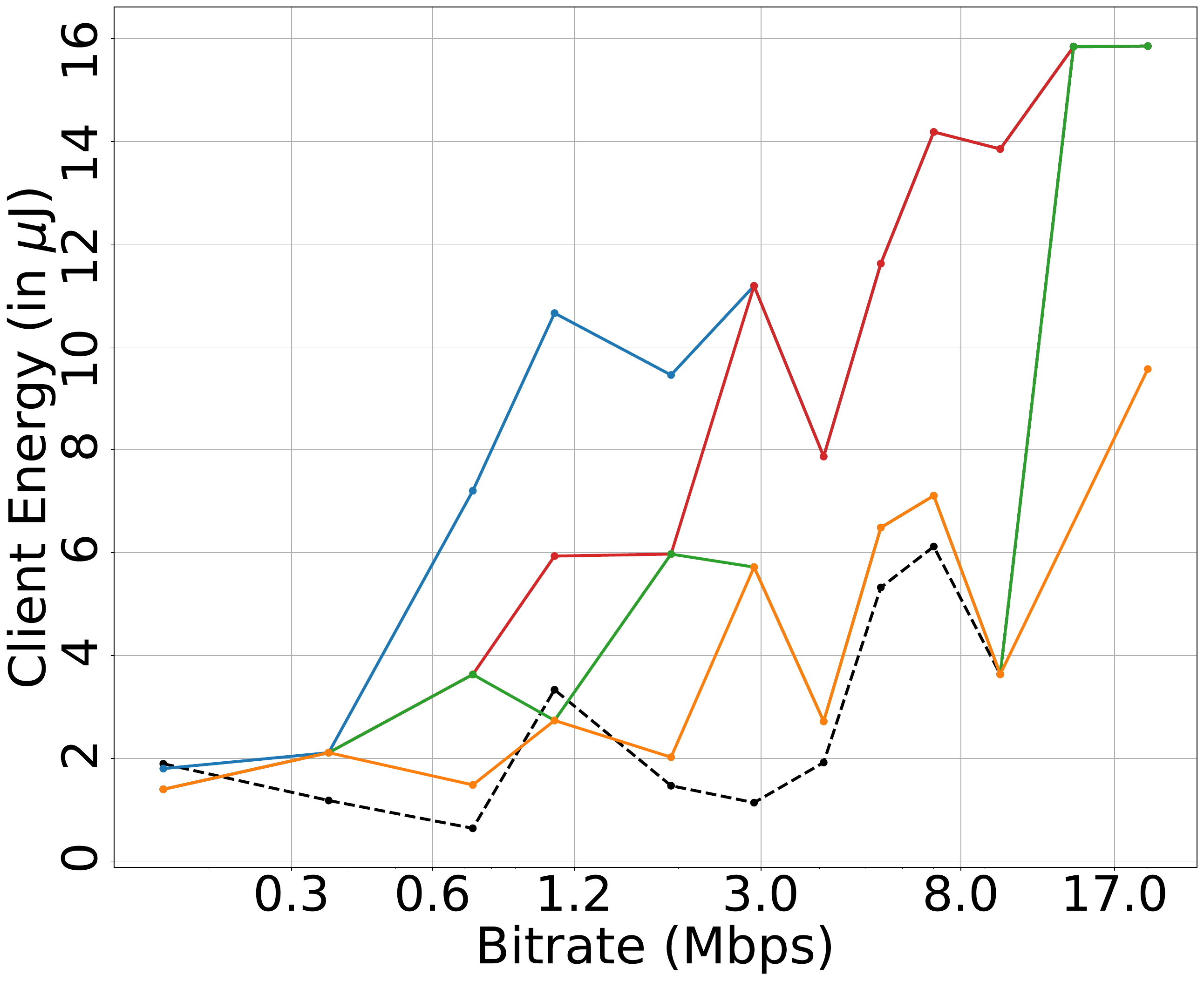} 
\caption{\emph{CourtYard} (\EY:103.08, \h:12.96, \LY:27.07)}
\end{subfigure}
\caption{Rate-quality, and rate-energy curves for the representative HDR sequences.}
\vspace{-0.8em}
\label{fig:rd_res}
\end{figure}

Fig.~\ref{fig:rd_res} illustrates 2D rate-quality and rate-energy behavior for representative HDR sequences. While configurations that maximize perceptual quality often incur higher client-side decoding energy, other operating points achieve comparable HDRVMAF at substantially lower energy cost. As a result, several configurations that appear attractive in the traditional rate–quality space become dominated when decoding energy is incorporated into the optimization objective. 

\subsubsection{Impact of Energy Weighting}
\amr{Table~\ref{tab:res_cons} reports the average performance of the evaluated bitrate ladders for different energy-weighting factors $\alpha_D$. At $\alpha_D = 0$, the optimizer prioritizes HDRVMAF and therefore favors high-resolution encodes. Increasing $\alpha_D$ shifts the selected representations toward lower-resolution, more energy-efficient operating points, often 1080p or 720p. As shown in Table~\ref{tab:res_cons}, this shift reduces client-side processing energy while maintaining substantial bitrate savings, confirming that energy-aware optimization significantly influences bitrate-ladder design.}

\subsubsection{Client Energy vs. Perceptual Quality}
Fig.~\ref{fig:rd_intro} and Fig.~\ref{fig:rd_res} show that \rev{client‑side processing energy} increases steeply with bitrate, even as HDRVMAF improvements plateau. Across multiple sequences, lower resolutions achieve nearly the same perceptual quality as higher resolutions but at substantially lower decoding and upscaling costs, making them an attractive operating point under energy constraints. 

\begin{table}[t]
\caption{The average results of the bitrate ladders compared to \emph{Default}. Arrows indicate the best performance trends.}
\centering
\resizebox{0.959\columnwidth}{!}{
\begin{tabular}{l||c|c|c|c|c}
\specialrule{.12em}{.05em}{.05em}
\specialrule{.12em}{.05em}{.05em}
Scheme & $\alpha_D$ &\BDRP & \BDRV & \BDRH & BDCE \\
& & [\%] $\downarrow$ & [\%] $\downarrow$ & [\%] $\downarrow$ $\uparrow$ & [\%] $\downarrow$ \\
\specialrule{.12em}{.05em}{.05em}
\specialrule{.12em}{.05em}{.05em}
DynResVMAF & 0 & -20.23 & -52.00 & -42.04 & 123.04 \\
\hline
DynResHDRVMAF & 0    & -20.15 & -50.12 & -54.34 & 27.15 \\
              & 0.02  & -16.80 & -42.32 & -47.60 & -10.30 \\
              & 0.04  & -15.95 & -41.84 & -47.41 & -11.86 \\
              & 0.16 & -11.20 & -36.51 & -42.21 & -33.40 \\
              & 0.32 &  -6.95 & -28.62 & -29.60 & -43.10 \\
\specialrule{.12em}{.05em}{.05em}
\specialrule{.12em}{.05em}{.05em}
\end{tabular}
}
\vspace{-0.75em}
\label{tab:res_cons}
\end{table}

\subsubsection{Content-Dependent Behavior}
Content characteristics strongly influence the optimal operating points (\cf~Fig.~\ref{fig:rd_res}). The representative sequences span distinct regions of the content feature space: \emph{NightStreet} exhibits lower spatial texture (\EY: 54.99) and motion (\h: 10.23) but higher brightness (\LY: 81.49), whereas \emph{CourtYard} contains richer spatial texture (\EY: 103.08) and motion (\h: 12.96) with substantially lower brightness (\LY: 27.07)~\cite{vca_ref}. Consequently, the more detailed and dynamic \emph{CourtYard} benefits from higher-resolution encodes to preserve perceptual quality, while the smoother \emph{NightStreet} produces flatter HDRVMAF curves, allowing lower-resolution encodes to achieve comparable quality with lower decoding energy. This variability indicates that a single, content‑agnostic ladder is insufficient and motivates per‑content, energy‑aware adaptation strategies.


\section{Conclusion}
This paper introduced a rate--quality--energy convex-hull framework for HDR video streaming that jointly models bitrate, HDR-aware perceptual quality, and client-side processing energy from decoding and upscaling. By evaluating candidate representations in this three-dimensional space, the framework reveals tradeoffs overlooked by traditional rate--quality optimization. Experiments on twelve 4K HDR sequences showed that many high-resolution encodes that appear optimal in the R--Q domain become dominated once energy is considered, with HDR-aware selections achieving substantial bitrate reductions, including \BDRH~values as low as \SI{-54.34}{\percent} and \BDRV~as low as \SI{-50.12}{\percent}, while maintaining equivalent perceptual quality. Energy-aware optimization also significantly reduces relative client-side energy cost, with BDCE improving from \SI{27.15}{\percent} at $\alpha_D=0$ to \SI{-43.10}{\percent} for larger energy weights. Consequently, increasing $\alpha_D$ shifts optimal operating points toward lower-resolution representations, reflecting HDRVMAF saturation at high resolutions and rapidly increasing processing energy. The proposed framework therefore enables device-aware HDR streaming through a tunable quality--energy tradeoff controlled by $\alpha_D$.

Future work includes adapting $\alpha_D$ to real-time device conditions, validating the framework on hardware-accelerated mobile platforms, and integrating it into DASH-based streaming systems for practical energy-efficient HDR delivery.
\balance
\newpage
\balance
\bibliography{references.bib}{}
\bibliographystyle{IEEEtran}
\balance
\end{document}